\documentclass[aps,prl,
   amsfonts,
   amssymb,
   amsmath,
   twocolumn,
   superscriptaddress,
   showpacs,
   preprintnumbers,
   ]{revtex4-1}
\usepackage{color}
\usepackage{graphicx}
\usepackage[pdftex,colorlinks=true,linkcolor=red,citecolor=blue]{hyperref}
\usepackage{time}
\newcommand{\coupling}{\lambda_A}          

\newcommand{\rd}{\mathrm{d}}               
\newcommand{\bW}{\mathbf{W}}

\newcommand{\bj}{\mathbf{j}}
\newcommand{\bk}{\mathbf{k}}

\newcommand{\bq}{\mathbf{q}}
\newcommand{\br}{\mathbf{r}}

\newcommand{\bnabla}{\boldsymbol{\nabla}}  
\newcommand{\calA}{\mathcal{A}}

\newcommand{\calD}{\mathcal{D}}

\newcommand{\calG}{\mathcal{G}}

\newcommand{\calL}{\mathcal{L}}
\newcommand{\calM}{\mathcal{M}}
\newcommand{\calN}{\mathcal{N}}

\newcommand{\calS}{\mathcal{S}}

\newcommand{\Diag}[1]{\mathrm{diag}( \, #1 \, )}   
\newcommand{\Set}[1]{ \bigl ( \, #1 \, \bigr )}    
\newcommand{\Expect}[1]
   {\ensuremath{\langle \, #1 \,  \rangle}}
\newcommand{\Comm}[2]
   {\ensuremath{[ \, #1, #2 \, ]}}
\newcommand{\AntiComm}[2]
   {\ensuremath{\{ \, #1, #2 \, \}}}
\newcommand{\Tr}[1]{\mathrm{Tr} [ \, #1 \, ]}      
\newcommand{\Exp}[1]{\exp [ \, #1 \, ]}            
\newcommand{\Ln}[1]{\ln [ \, #1 \, ]}              
\newcommand{\Det}[1]{\det [ \, #1 \, ]}            
\newcommand{\Intk}{\int \!\! %
   \frac{ \mathrm{d}^3 k }{ (2\pi)^3 } }           
\begin{document}
%
%
\preprint{LA-UR 11-02991; INT-PUB-11-018} 
\title[Anomalon]
   {Composite-Field Goldstone States and Higgs Mechanism in Dilute Bose Gases
   }
\author{Fred Cooper}
\affiliation{
   Los Alamos National Laboratory,
   Los Alamos, NM 87545}
\affiliation{Santa Fe Institute,
   Santa Fe, NM 87501}

\author{Chih-Chun Chien}
\affiliation{
   Los Alamos National Laboratory,
   Los Alamos, NM 87545}
   
\author{Bogdan Mihaila}
\affiliation{
   Los Alamos National Laboratory,
   Los Alamos, NM 87545}

\author{John F. Dawson}
\affiliation{Department of Physics,
   University of New Hampshire,
   Durham, NH 03824}

\author{E. Timmermans}
\affiliation{
   Los Alamos National Laboratory,
   Los Alamos, NM 87545}
%
%
%
%
\begin{abstract}
We show that a composite-field (diatom) Goldstone state is expected in a dilute Bose gas for temperatures between the Bose gas critical temperature where the atom Bose-Einstein condensate appears 
and the temperature where superfluidity sets in. The presence of superfluidity is tied to the existence of a U(1) charge-two diatom condensate in the system. 
By promoting the global  U(1) symmetry of the theory to a gauge symmetry,
we find that the mass of the gauge particle generated through the Anderson-Higgs mechanism is related to the superfluid density via the Meissner effect 
and the superfluid density is related to the square of the anomalous density in the Bose system. 
\end{abstract}
%
%
%
%
\maketitle
%
%

Ultracold atomic gases provide a fertile testbed for many-body theory. Tabletop experiments can be designed to take advantage of the ability of tuning the strength of the inter-particle interactions and provide new insights in the role played by correlations in diverse physics problems such as high-T$_c$ superconductors and the equation of state for neutron matter. To this growing list of applications, we would like to add the Higgs mechanism, which is of great interest because of the ongoing experiments at the Large Hadron Collider and their potential implications to the physics of the Standard Model of elementary particles. In addition, the Higgs mechanism in ultracold atomic gases is particularly intriguing because recent advent of synthetic magnetic fields allow for the study of electromagnetic-like effects in systems of neutral atoms~\cite{ref:spielman09,*ref:spielman11}.

Recently we reformulated the theory of dilute Bose gases in terms of normal and anomalous auxiliary-field (AF) densities~\cite{PhysRevLett.105.240402}. In the leading order of the AF loop expansion (LOAF) we find two critical temperatures in the phase diagram: the critical temperature~$T_c$, where the atom Bose-Einstein condensate (BEC) appears first, and a temperature~$T^{\star} > T_c$, which indicates the onset of superfluidity and the appearance of a diatom condensate in the system.
In this paper we will show that between $T_c$ and $T^{\star}$ the system supports zero-energy and zero-momentum excitations, which correspond to composite-field (diatom) Goldstone states.  
Introducing a formal U(1) gauge vector meson into the system, we find that the mass of the gauge particle generated through the Anderson-Higgs mechanism~\cite{PhysRev.117.648,*PhysRev.130.439,*PhysRevLett.13.321,*PhysRevLett.13.585,*r:Higgs:1964zr,*PhysRevLett.13.508} can be related to the superfluid density via the Meissner effect. 

%
%
In the LOAF theory of dilute Bose gases~\cite{PhysRevLett.105.240402,PhysRevA.Bose:2011} 
the classical action is given by $\calS[\Phi] = \int \rd^4 x \, \calL[\Phi]$, where
\begin{gather}
   \calL[\Phi]
   =
   \frac{1}{2 \lambda}
   \bigl [ \,
      \chi^2(x) - | A(x) |^2 \,
   \bigr ] 
   -
   \sqrt{2} \, \chi(x) \, | \phi(x) |^2 
   \label{e:S-I}\\
   +
   \bigl [ \,
      A^{\ast}(x) [ \phi(x) ]^2 
      +
      A(x) [ \phi^{\ast}(x) ]^2 \,
   \bigr ]
   \notag \\
   +
   \frac{1}{2} 
   \bigl [ \,
      \phi^{\ast}(x) \, h \, \phi(x) + \phi(x) \, h^{\ast} \, \phi^{\ast}(x) \,
   \bigr ] \>,
   \notag 
\end{gather}
with $h = i \hbar \, \partial_t + \gamma \nabla^2 + \mu$ and $\gamma = \hbar^2 / (2m)$.  
The coupling constant, $\lambda$, is related to $a_0$, the s-wave scattering wave~\cite{PhysRevA.Bose:2011}.
Here, $\chi(x)$ and $A(x)$ are real and complex fields related to $|\phi(x)|^2$ and $\phi^2(x)$, respectively, and $\Phi^{\alpha} = \Set{\phi,\phi^{\ast},\chi,A,A^{\ast}}$ is a set of five fields.  
The Lagrangian density in Eq.~\eqref{e:S-I} possesses a global U(1) symmetry under the transformations
\begin{equation}
   \phi \rightarrow  e^{i \Lambda} \phi \>,
   \qquad
   \label{e:Usym} 
   A \rightarrow  e^{ 2i \Lambda} A \>,
\end{equation}
Adding sources terms $J^{\alpha}(x) \equiv \Set{ j,j^{\ast},s,S,S^{\ast}}$ to the Lagrangian density, the generating functional $W[J]$ of connected graphs is obtained from the path integral by~\cite{r:Negele:1988fk}
\begin{gather}
   Z[J]
   =
   e^{i W[J] / \hbar}
   = 
   \calN
   \int \calD \Phi \>
   e^{ i \calS'[\Phi,J] / \hbar } \>,
   \label{pathint} \\
   \calS'[\Phi,J] 
   = 
   \calS[\Phi] 
   + 
   \int \rd^4 x \
   J^{\alpha}(x) \, \Phi_{\alpha}(x) \>.
   \notag
\end{gather} 
The generator of one-particle irreducible (1-PI) graphs, $\Gamma[\Phi]$, is obtained by a Legendre transformation from the classical currents $J^{\alpha}(x)$ to the classical fields $\Phi_{\alpha}(x)$ via 
\begin{equation}
   \Gamma[\Phi] 
   = 
   \int \rd^4 x \ J^{\alpha}(x) \, \Phi_{\alpha}(x)
   - 
   W[J] \>,
\end{equation}
The equations of motion and the inverse propagator are 
\begin{equation}\label{e:eoms}
   \frac{\delta \Gamma[\Phi]}{\delta \Phi_{\alpha}(x)} 
   = 
   J_{\alpha}(x) \>,
   \quad
   \frac{\delta^2\Gamma[\Phi]}
        {\delta \Phi^\alpha(x) \, \delta \Phi^{\gamma}(x')} 
   = 
   \calG^{-1}_{\alpha\gamma}(x,x') \>.
\end{equation}
The path integration over the $\phi(x)$ 
fields is done exactly and the integral over the fields $\chi(x)$, $A(x)$ and $A^{\ast}(x)$ is approximated by steepest descent.  We obtain
\begin{align*}
   \Gamma[\Phi]
   = &
   \frac{1}{2} \!\iint \! \rd^4 x \, \rd^4 x' 
   \phi_a^{\ast}(x) \, G^{-1}_{ab}[\chi,A](x,x') \, \phi_b^{\phantom\ast}(x')
   \notag \\ 
   &
   \!\! - \!\!
   \int \!\! \rd^4 x \,
   \Bigl \{ 
      \frac{\chi^2 - | A |^2}{2\lambda} \,
      - 
      \frac{\hbar}{2i}  
      \Tr{ \Ln{ G^{-1}[\chi,A](x,x) } } 
   \Bigr \} \>.
\end{align*}
Here, $G^{-1}_{ab}$ represents the $\{1,2\}$ sector of $\calG^{-1}_{\alpha \beta}$, i.e.
\begin{align}
   &G^{-1}_{ab}[\chi,A](x,x')
   \label{e:G0invdef} 
   \\ \notag & 
   = 
   \delta(x,x') \! 
   \begin{pmatrix}
      -
      i \hbar \, \partial_t
      -
      \gamma \,  \nabla^2
      + \chi^\prime
      & - A \\
      - A^{\ast} & 
      i \hbar \, \partial_t
      -
      \gamma \, \nabla^2
      + \chi^\prime
   \end{pmatrix} \! ,
\end{align}
where we introduced the notation $\chi^\prime =  \sqrt{2} \, \chi  -  \mu $.  
The finite-temperature partition function for the system is given by the same path integral~\eqref{pathint} in the Matsubara imaginary time ($t \mapsto - i  \hbar\tau $) formalism~\cite{r:Negele:1988fk}.
In the imaginary-time formalism  and for constant values of $\chi^\prime$ and $A$, the Green function $G_{ab}[\chi,A](x,x')$ depends only on $x-x'$. In Fourier space, we have
\begin{equation}\label{e:calGexpand}
   \calG_{\alpha \beta}(x)
   = \!\!
   \Intk \sum_{n}
   \tilde{G}_{\alpha \beta}(\bk,n) \, 
   e^{i [ \,\bk \cdot \br - \omega_n t \,]} \>,
\end{equation}
where $\omega_n = 2 \pi n/T$ are the Matsubara frequencies in $k_B=1$ units.  From Eq.~\eqref{e:G0invdef}, we find
\begin{align}  
   \tilde{G}_{11}(\bk,n)
   &=
   \tilde{G}_{22}^{\ast}(\bk,n)
   =
   \frac{\epsilon_{k} + \chi^\prime + i \omega_{n}}
        {\omega_{n}^{2}+\omega_{k}^2} \>,
   \label{e:Green} \\
   \tilde{G}_{12}(\bk,n)
   &=
   \tilde{G}_{21}^{\ast}(\bk,n)
   =
      \frac{A}
        {\omega_{n}^{2}+\omega_{k}^2} \>,
   \notag
\end{align}
with $\epsilon_k = \gamma k^2$ and the dispersion relation
\begin{equation}\label{eq:dispersion}
    \omega_k^2 = ( \epsilon_k + \chi^\prime )^2 - |A|^2
    \>,
\end{equation}
The effective potential at finite temperature is\footnote{Alternatively, one can start with the hamiltonian density describing a dilute Bose gas, given by 
$H$=$ \int d^3x \bigl [  \phi^\ast(x,t) \bigl( - \gamma \nabla^2 \bigr )  \phi(x,t)  + (\lambda/4) ( \phi^\ast \phi)^2   \bigr ]
$.
The Hamiltonian is invariant under the $U(1)$ transformation $\phi \rightarrow e^{i\Lambda} \phi$ and its complex conjugate. 
In the functional approach to the imaginary time formalism of Matsubara, the grand canonical partition function, $Z$, is given by a path integral, 
$
Z $=$ \rm{Tr} \ e^{-\beta(H-\mu N) } $=$ \int {\cal D} \phi^\ast \ {\cal D \phi} \ e^{-S[\phi^\ast,\phi]}
$,
where the imaginary time action $S$ is given
$
S[\phi^\ast, \phi] = \int_0^\beta  \!\! d\tau \, d^3x 
\bigl [ \phi^\ast(x,t) \bigl ( \partial_\tau - \mu \bigr ) \phi(x,t) + H(\phi^\ast,\phi) \bigr ] \>.
$ 
Performing the Hubbard-Stratonovich transformation and introducing external sources,  one can integrate over the $\phi$ fields exactly and perform the remaining integral by the Laplace method leading to a loop expansion in the composite field propagator~\cite{PhysRevLett.105.240402}. Legendre transforming from the sources to the expectation values of the fields, one has (for homogeneous fields)  that in lowest order in this expansion the Gibbs free energy,  $F$=$- (T / V) \bigl ( \ln Z - \int d^3x \, j_i \phi_i \bigr )$, i.e.  the finite temperature effective potential,~$V_\text{eff}$, in Eq.~\eqref{e:Veff} }
\begin{align}  
   &V_{\text{eff}}[\Phi]
   =
   \chi^\prime | \phi |^2
   -
   \frac{A^{\ast} \, \phi^2 }{2} 
   -
   \frac{A \, \phi^{\ast\,2} }{2}
   -
   \frac{ (\chi^\prime + \mu)^2 }{4\lambda} 
   \label{e:Veff} \\
   & \quad
   +
   \frac{|A|^2}{2\lambda}
   +
   \Intk
   \Bigl [ \,
      \frac{\omega_k}{2}
      +
      T \,  \ln \bigl ( 1 - e^{-\beta \omega_k} \bigr ) \,
   \Bigr ] \>,
   \notag 
\end{align}
and the particle density is given by
$   \rho
   =
   - \,
  \partial V_{\text{eff}} / \partial \mu
   = 
   (\chi^\prime + \mu)/(2 \lambda) 
$. 
The last term in Eq.~\eqref{e:Veff} corresponds to a term $ {\rm Tr~ln} \, G^{-1}$ in the LOAF effective action~\cite{PhysRevLett.105.240402}. 
Minimizing the effective potential, $V_\text{eff}$, with respect to $\phi^\ast$  gives the minimum condition
\begin{equation}
    \chi^\prime \phi - A \,\phi^{\ast} = 0 \>.
    \label{e:phi0}
\end{equation}
Because of the gauge freedom, we can choose $\phi$ to be real at the
minimum which means that for the broken symmetry case ($\phi \ne 0$),
$A = \chi^\prime$ is real at the minimum, and leads to the Bogoliubov
spectrum at weak coupling~\cite{PhysRevLett.105.240402}.  Minimizing
the effective potential with respect to $\chi^\prime $ and $A$ leads to
the self-consistent equations:
\begin{align}
   \frac{A}{\lambda}
   &=
   \rho_{0}
   +
   A \Intk \,
      \frac{1 + 2 n(\omega_{k}/T)}{2\omega_{k}} \,
   \>,
   \label{e:EOS-I} \\
   \rho
   &=
   \rho_{0}
   +
   \Intk \,
      \frac{\epsilon_{k} + \chi^\prime}{2\omega_{k}} \,
      [ \, 1+2n(\omega_{k}/T) \, ] \,
   \>.
   \label{e:EOS-II}
\end{align}
Here $\rho_0 = |\phi|^2$ and $n(x) = (e^{x}-1)^{-1}$ is the Bose distribution.  Eq.~\eqref{e:Veff} is regularized as shown in Ref.~\onlinecite{PhysRevA.Bose:2011}.

\begin{figure}
   \includegraphics[width=\columnwidth]{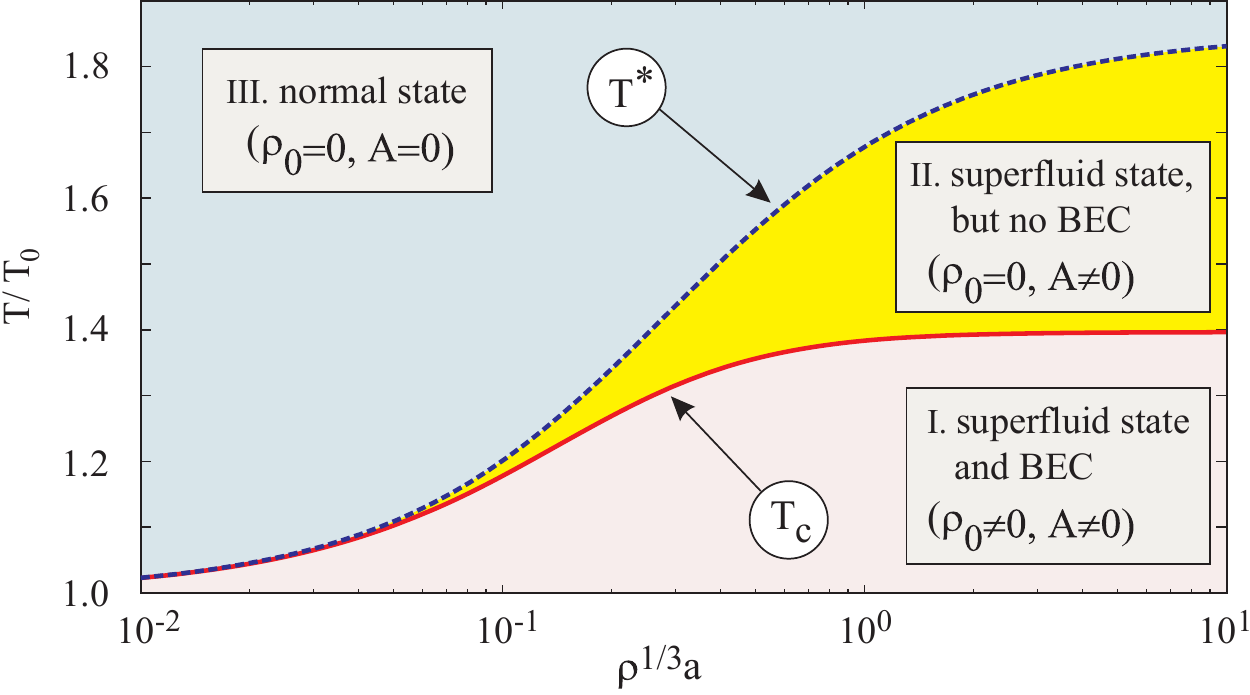}
   \caption{\label{fig:PD}(Color online) 
   The LOAF phase diagram.  
   }
\end{figure}

%
%
The phase diagram in the LOAF approximation is depicted in Fig.~\ref{fig:PD} as a function of the strength of the inter-particle interaction, which is characterized by the dimensionless parameter $\rho^{1/3}a$, where $a_0$ is the s-wave scattering length. 
We notice the presence of three distinct regions, corresponding to the
values of the three LOAF parameters, the usual (atom) BEC condensate density,
$\rho_0$, and the normal and anomalous auxiliary fields, $\chi^\prime$
and $A$: (I)~$0 < T < T_c$, with $\rho_0 \ne 0$ and $\chi^\prime = A >
0$, (II)~$T_c < T < T^\star$, with $\rho_0 = 0$ and $\chi^\prime > A >
0$, and (III)~$T > T^\star$, with $\rho_0 = A = 0$ and $\chi^\prime >
0$. Here, the critical temperature~$T_c$ corresponds to the emergence
of the atom BEC condensate, whereas the temperature~$T^\star$  is related
to the onset of superfluidity in the system and the emergence of a diatom condensate,~$A$.  
The anomalous auxiliary field~$A$ represents a second order parameter in the LOAF theory.  Region~I in the phase diagram features both an atom  BEC condensate and a superfluid state.  In region~II the atom BEC condensate is no longer present, but part of the system is still in a superfluid state, whereas in region~III the entire system is in the normal state.  In the noninteracting limit, $T_c$ and $T^\star$ are the same.  As the interaction strength increases, the temperature range for which the superfluid is present in the absence of the atom BEC condensates expands. LOAF predicts $\sim20$\% temperature range $T^\star - T_c$ relative to $T_c$  for a dimensionless parameter value, $\rho^{1/3}a =1$.   
In regions~I and~II fields carrying  $U(1)$ charge are nonzero, which leads to spontaneous breaking of the $U(1)$ charge and the existence of Goldstone modes.  These Goldstone modes are essential for the existence of superfluidity according to the  Josephson relationship~\cite{r:josephson66}.

%
%

\begin{figure}
   \includegraphics[width=\columnwidth]{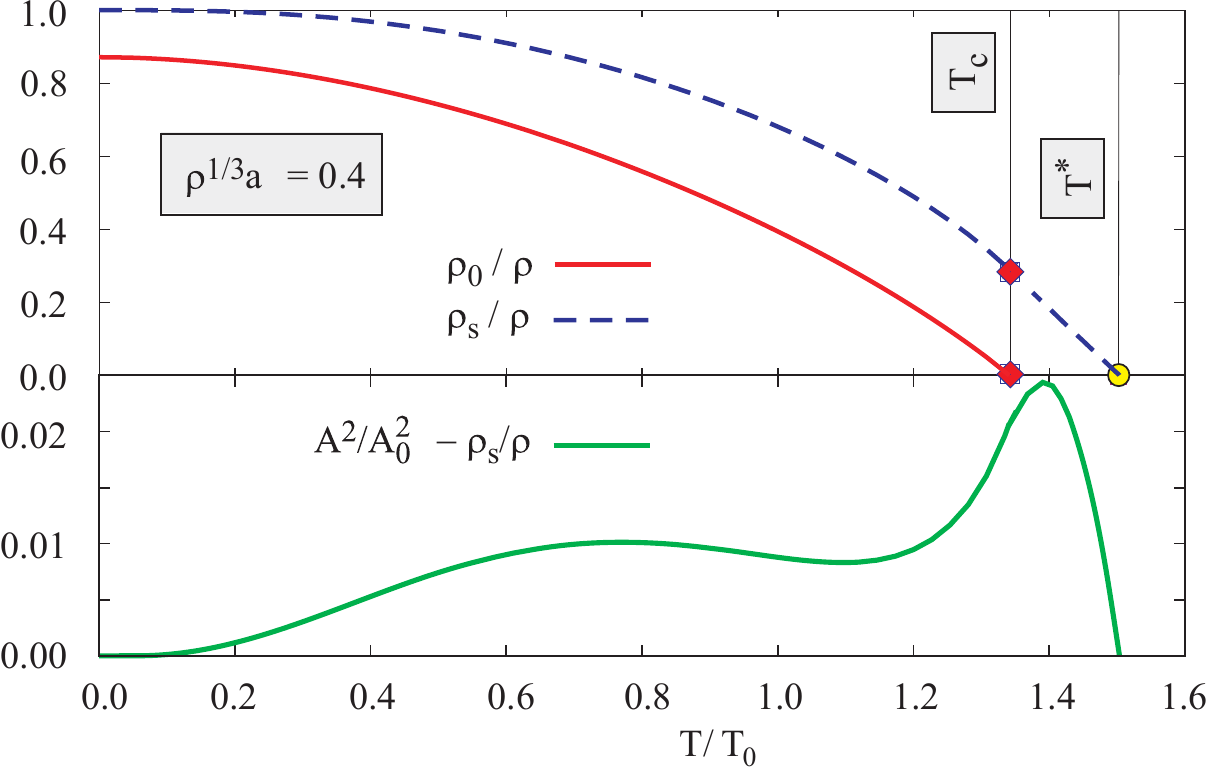}
   \caption{\label{fig:SF}(Color online) 
   (top) Comparison of the atom BEC condensate density,  $\rho_{0}$, and the superfluid density, $\rho_s$,
   for $\rho^{1/3}a_{0}=0.4$.
   (bottom) Comparison of $A^2$ and~$\rho_s$. Here $A_0 \equiv A(T=0)$.
   }
\end{figure}

The superfluid state in the Bose gas is discussed using Landau's phenomenological two-fluid model~\cite{r:Landau:1941ly,*r:Landau:1947}. We define the normal-state density
\begin{equation}\label{eq:rhon}
   \rho_{n}
   =
   \frac{1}{3\pi^2}\int_{0}^{\infty}dk \, k^2 \ \epsilon_k \,
   \Bigl [ \,
      -\frac{\partial n(\omega_k)}{\partial \omega_k} \,
   \Bigr ] \>,
\end{equation}
where $\omega_k$ is the quasi-particle energy given by the LOAF dispersion relation~\eqref{eq:dispersion}.
Correspondingly, the superfluid density is defined as $\rho_{s}=\rho-\rho_{n}$. In Fig.~\ref{fig:SF}, we illustrate the temperature dependence of the atom BEC condensate, $\rho_0$, and the superfluid density, $\rho_s$, relative to the system density,~$\rho$, for an interaction strength $\rho^{1/3}a = 0.4$. As advertised, the onset of superfluidity in the system occurs at $T^\star$, and this temperature is different from the $T_c$, the emergence temperature of the atom BEC condensate.

If we promote the U(1) symmetry to a gauge symmetry by setting $\Lambda \rightarrow g \Lambda(x)/\hbar$ where $g$ is the U(1) charge, in order to preserve the symmetry we introduce a gauge field~$W_{\mu}(x) = \Set{W_0(x),\bW(x)}$ analogous to the weak interaction vector gauge field.  
Then, the boson fields, $\phi$, and the anomalous auxiliary field, $A$, carry one and two units of U(1) charge, respectively. The LOAF approximation predicts that the superfluid state is accompanied by a Meissner effect 
in the presence of a weak vector potential, $\bW(x)$.  Following the standard derivation of the supercurrent in the Bardeen-Cooper-Schrieffer (BCS) theory (see e.g.\ Secs.~51-52 in Ref.~\onlinecite{r:Fetter:1971fk}) and adapting it to the case of a Bose gas of particles, one obtains the supercurrent in the low-momentum limit, as
\begin{equation}\label{eq:superC}
    \bj_s(\bq) = - \rho_{s}  \, \frac{g^2}{mc} \bW(\bf q)
    \>.
\end{equation}
Unlike in the BCS case, in LOAF the superfluid density $\rho_s$ in Eq.~\eqref{eq:superC} is the same as that obtained in the Landau two-fluid model described above. 
Characteristic signatures of superfluidity in Bose atom gases, such as
dissipationless flow~\cite{PhysRevLett.83.2502,*PhysRevLett.99.260401}
and formation of quantized vortices in rotating
gases~\cite{PhysRevLett.84.806},  have already been observed.
Superfluidity can also be measured directly by analyzing the ``scissors'' mode~\cite{PhysRevLett.84.2056} with the atomic cloud oscillating with respect to the symmetry axis of the confining potential.  In nuclear physics this mode was observed in heavy deformed nuclei~\cite{Bohle:1984,*Richter:1995}.  
Recently, Cooper and Hadzibabic proposed using
a vector potential generated by optical beams with nonzero angular momenta to simulate uniform
rotation  of the atomic gas~\cite{PhysRevLett.104.030401}. The induced change in angular momentum can be measured spectroscopically and provides a determination of the superfluid fraction.

%
%


The Goldstone theorem~\cite{r:Goldstone:1961kx} states that when a continuous symmetry, such as U(1), is spontaneously broken, then, necessarily, new massless scalar states appear in the excitation spectrum related to the order parameter. 
In LOAF, the composite-field Goldstone theorem for the auxiliary field~$A$ gives rise to a massless scalar when $T_c < T < T^\star$.  
To derive the Goldstone theorem, we 
consider Noether's theorem for the U(1) transformation~\eqref{e:Usym} on the fields $\Phi$.  We 
calculate the change in the Lagrangian density, $\calL' = \calL + J_\alpha \Phi^\alpha$, 
under the infinitesimal change, $\Phi^\alpha(x)  \rightarrow \Phi^\alpha(x) + \delta  \Phi^\alpha(x)$, 
with~(see e.g. Sec. 19 in Ref.~\onlinecite{r:WeinbergII})
\begin{equation}
    \delta  \Phi^\alpha(x) = \frac{i}{\hbar} \ \epsilon \  g^\alpha{}_\beta \, \Phi^\beta(x)
    \>,
\end{equation}
where we introduced the U(1) charge matrix,
$
   g^\alpha{}_\beta
   =
   g \ \Diag{ 1,-1,0,2,-2 } 
$, for the fields $\Phi^\alpha = \{ \phi, \phi^\ast, \chi, A, A^\ast \}$ corresponding to $\alpha=1\cdots 5$. 
The complex fields $\phi(x)$ and $A(x)$ carry U(1) charges $g$ and $2g$, respectively. The real field~$\chi$ is U(1) neutral. 
From
\begin{equation*} 
   \delta \calL' 
   = 
   \frac{\delta \calL'}{\delta \Phi_\alpha } \delta \Phi_\alpha 
   + 
   \frac{\delta \calL'}{\delta \, \partial_\mu  \Phi_\alpha} \, \delta \, \partial_\mu \Phi_\alpha  
   = 
   \partial_\mu 
   \Bigl ( 
      \frac{\delta \calL'}{\delta \, \partial_\mu  \Phi_\alpha}  \, \delta \Phi_\alpha 
   \Bigr ) \>,
\end{equation*}
we obtain
\begin{equation}\label{e:curcons}
   \frac{\partial \rho_0(x)}{\partial t}
   +
   \bnabla \cdot \bj_0(x)
   =
   (i/\hbar) \,
   J_{\alpha} \, g^\alpha{}_\beta \, \Phi^{\beta} \>,   
\end{equation}
where
\begin{align}
   \rho_0(x) 
   &= 
   g | \phi(x) |^2 \>,
   \label{e:rho0def} \\
   \bj_0(x) 
   &= 
   \frac{ g \hbar}{2 m i } \, 
   \bigl [ \, 
      \phi^{\ast}(x) \bnabla \phi(x) 
      -  
      \phi(x) \, \bnabla \phi^{\ast}(x) \, 
   \bigr ] \>.
   \notag
\end{align}
Eq.~\eqref{e:curcons} is a classical result, representing the U(1) current conservation in the absence of external sources. 
We multiply \eqref{e:curcons} by $\Exp{- S[\Phi,J]/\hbar}$, divide by~$Z$, and integrate over the fields $\Phi$ to derive the quantum version.
Next, we substitute $ J_\alpha(x) = \delta \Gamma[\Phi] / \delta
\Phi^{\alpha}(x)$ from Eq.~\eqref{e:eoms} in the resulting equation, where $\Gamma$ is the
Legendre transform of $\ln Z$ and generates one-particle irreducible
graphs. We integrate over $\rd^4 x$ and discard  the surface terms to
obtain the functional equation~\cite{r:WeinbergII}
\begin{equation}\label{e:dcurcons}
   \int \rd^4 x \
   \frac{\delta \Gamma[\Phi]}{\delta \Phi^{\gamma}(x)} \, 
   g^\gamma{}_\beta \, \Phi^{\beta}(x)
   =
   0 \>.
\end{equation}
Differentiating \eqref{e:dcurcons} with respect to $\Phi^{\alpha}(x)$ gives
\begin{align}
   &\int \rd^4 x' \
   \frac{\delta^2 \Gamma[\Phi]}
        {\delta \Phi^{\alpha}(x) \, \delta \Phi^{\gamma}(x')} \, 
   g^\gamma{}_\beta \, \Phi^{\beta}(x')
   \label{e:secondderv} \\
   &=
   \int \rd^4 x' \
   \calG_{\alpha \gamma}^{-1}(x,x') \,
   g^\gamma{}_\beta \, \Phi^{\beta}(x')
   =
   - \, J_{\gamma}(x) \ g^\gamma{}_\alpha \>.
   \notag
\end{align}
For constant fields 
and in the absence of sources, Eq.~\eqref{e:secondderv} gives
$
   \calM_{\alpha \beta} \,\Phi^{\beta}
   =
   0 
$,
where $\calM_{\alpha \beta}$ is a 4$\times$4 matrix with indices $\{1,2,4,5\}$ and
\begin{align}
   \calM_{\alpha \beta}
   & =
   \int \rd^4 x' \
   \calG^{-1}_{\alpha \gamma}(x,x') \, g^\gamma{}_\beta
   = \
   \tilde
   \calG^{-1}_{\alpha \gamma}(0,0) \, g^\gamma{}_\beta 
   \>.
\end{align}
%
The Goldstone theorem corresponds to $\Det{\calM} = 0$ and implies the presence of a pole in the propagator at zero-energy and zero-momentum transfer~\cite{r:WeinbergII}. 
This result is preserved order-by-order in the auxiliary-field (AF) loop expansion, because the Ward identities are preserved order-by-order.
In the LOAF approximation, the $\phi$ propagator is leading order, but the mixing between the  $\phi$ and $A$ propagators only arises at  next-to-leading order (NLO) and is proportional to~$\phi$.
Hence, $\Det{\calM}$ is given by the product of the determinants of the $\{1,2\}$ and $\{4,5\}$ diagonal blocks, respectively, plus NLO corrections. The latter are dropped in LOAF. Then, the zero-determinant condition is satisfied if the determinant of either diagonal block vanishes.
Furthermore, at the minimum of the effective potential, the fields can be taken to be real, and the two determinant conditions are equivalent with
\begin{eqnarray}
  \label{e:Goldstone_phi}
   \bigl [ \tilde{\calG}^{-1}_{11}(0,0)
   & - 
   \tilde{\calG}^{-1}_{1 2}(0,0) \bigr ] \, \phi
   = & 
   0 \>,
   \\
   \label{e:Goldstone_A}
   \bigl [ \tilde{\calG}^{-1}_{4,4}(0,0)
   & - 
   \tilde{\calG}^{-1}_{5,4}(0,0) \bigr ] \, A
   = & 
   0 \>.  
\end{eqnarray}
Nontrivial solutions of Eqs.~\eqref{e:Goldstone_phi} and \eqref{e:Goldstone_A} require the brackets to vanish. These are the  broken-symmetry Ward identities for the $\phi$ and $A$ propagators, repectively.

The Goldstone theorem for the atom BEC condensate corresponds to the case $\phi \neq 0$. Hence, Eq.~\eqref{e:Goldstone_phi} is consistent with the minimum condition~\eqref{e:phi0}, which is satisfied in region~I of the LOAF phase diagram: for $T<T_c$ with $\chi^\prime = A$. 
In LOAF we find that in general
\begin{align}
   &
   \tilde{\calG}^{-1}_{4,4}(0,0)
   -
   \tilde{\calG}^{-1}_{4,5}(0,0)
   \label{e:GoldLOAF} \\
   &
   =
   \frac{1}{2} \,
   \Bigl [ 
      \frac{1}{\lambda}
      -
      \int \frac{\rd^{3}k}{(2\pi)^3} \,
         \frac{1 + 2 n(\omega_{k}/T)}{2\omega_{k}} 
   \Bigr ]
   =
   \frac{\rho_0}{2 A} \>.
   \notag   
\end{align}
Here we used
\begin{align*}
   \tilde{\calG}^{-1}_{4,4}(0,0)
   &=
   \frac{1}{2\lambda}
   + 
   \frac{T}{2} \Intk \sum_n \, 
   \tilde{G}_{11}(\bk,n) \, \tilde{G}_{22}(\bk,n) \>,
  \\
   \tilde{\calG}^{-1}_{4,5}(0,0)
   &=
   \frac{T}{2} \Intk \sum_n \, 
   \tilde{G}_{12}(\bk,n) \, \tilde{G}_{12}(\bk,n) \>,
\end{align*}
with the Green functions $\tilde{G}_{ab}$ given in Eq.~\eqref{e:Green}. In region II ($T_c < T < T^\star$)  where  $\rho_0$ is zero
but $A \neq 0$, 
 Eq.~\eqref{e:GoldLOAF} is identically zero and we find a composite-field Goldstone theorem, 
corresponding to a zero energy and momentum excitation of the gas.

%
%

%
%
Next, we connect  the superfluidity and the  Higgs effect.
From the supercurrent~\eqref{eq:superC} and the Maxwell equation, 
$\bnabla \times ( \bnabla \times \bW ) = \bj_s$, we find the London equation~\cite{london:1935}
\begin{equation}\label{e:Beq}
   \nabla^2 \, \bW
   +
   \frac{c}{4\pi \, \lambda_L^2} \,
   \bW
   =
   0 \>,
\end{equation}
where $\lambda_L = \sqrt{ m c^2 / (4\pi \rho_s g^2) }$ is the London penetration depth for the $\bW$ field.
To relate $\rho_s$ to the fundamental quantities of our theory,  we note that the $\mathrm{Tr}\{ \ln G^{-1} \}$ term in the action leads to nonlocal temperature-dependent $n-\phi$ vertices that are polygons in the Green function, $G$. Following a similar approach in BCS theory~\cite{PhysRevLett.71.3202,*PhysRevB.55.15153}, we write a 
low-momentum Ginzburg-Landau effective theory for the $2g$-charge composite  field,~$\calA$.  For simplicity, we use the relativistic generalization of our theory  to make contact with the Anderson-Higgs mechanism~\cite{PhysRev.117.648,*PhysRev.130.439,*PhysRevLett.13.321,*PhysRevLett.13.585,*r:Higgs:1964zr,*PhysRevLett.13.508}.  By symmetry arguments, the relativistic effective field theory for the field~$\calA$ in the presence of a U(1) gauge field has the form
\begin{equation}
   \calL
   = 
   (D_\mu \calA )^{\ast} \, ( D^\mu \calA ) 
   - 
   \coupling \, \bigl ( |\calA|^2 - A^2 \bigr )^2
   - 
   \tfrac{1}{4} F_{\mu \nu} F^{\mu \nu} \>,
\end{equation}
with the covariant derivative, $D_\mu = \partial_{\mu} + 2 i g W_{\mu}$, and $F_{\mu \nu} = \partial_{\mu} W_{\nu} - \partial_{\nu} W_{\mu}$.  
Here, $\coupling \equiv \coupling(T)$ is the value of the four-point field interaction at zero momentum transfer and temperature~$T$, and $A\equiv A(T)$ is obtained by solving Eqs.~\eqref{e:EOS-I} and~\eqref{e:EOS-II} for $T < T^{\star}$.  Note that when $g \rightarrow 0$, this is the usual effective Lagrangian for charged scalars which exhibits the Goldstone theorem. 
 Using the gauge freedom, we expand the field about $A$ as 
\begin{equation}\label{e:expandA}
   \calA
   =
   A
   +
   \tfrac{1}{\sqrt 2} \, \bigl ( \calA_1 + i \calA_2 \bigr ) \>,
   \quad \Expect{\calA_1} = \Expect{\calA_2} = 0 
   \>,
\end{equation}
which leads to the Lagrangian of the form
\begin{align}
   \calL
   = \, &
   \tfrac{1}{2} (\partial_\mu \calA_1)^2
   -
   \tfrac{1}{2} \, \coupling \,  A^2 \, \calA_1^2
   \label{e:lagII} 
   \\ \notag
   & 
   -
   \tfrac{1}{4} \, F_{\mu\nu} F^{\mu\nu}
   +
   \tfrac{1}{2} 
   \bigl (  
   \partial_\mu \calA_2 + 2 \sqrt{2} \, g A \ W^{\mu} 
   \bigr )^2
   \ +
   \cdots
   \>.
\end{align}
Introducing a new field:
$
   W'_{\mu}
   =
   W_{\mu}
   +
   \partial_{\mu} \calA_2 /  (2 \sqrt{2} \, g A)
$, 
the Lagrangian \eqref{e:lagII} becomes
\begin{align}  
   \calL
   &=
   \tfrac{1}{2} (\partial_\mu \calA_1)^2
   -
   \tfrac{1}{2} \, \coupling \, A^2 \, \calA_1^2
   -
   \tfrac{1}{4} \, F'_{\mu\nu} F^{\prime\,\mu\nu}
   \label{e:lagIII} \\
   & \quad
   +
   ( 2 g A )^2 \, W'_{\mu} W'{}^{\mu}
   +
   \cdots
   \>.
   \notag
\end{align}
where $F'_{\mu\nu} = \partial_{\mu} W'_{\nu} - \partial_{\nu} W'_{\mu}$.
Hence, the field~$\calA_1$ has a composite-field Higgs mass, $M_H^2 = \coupling \, A^2$, whereas the effective mass of the gauge field~$W'_\mu$ is $M_W^2 = ( 2 g \, A )^2$. The latter is identified as
\begin{equation}
    M_W^2 = ( 2 g \, A )^2 \ \longrightarrow \ \rho_s \, \frac{g^2}{mc^2}  \label{mass}
    \>.
\end{equation}
This implies that $A^2$ is a measure of the superfluid density,~$\rho_s$.  In Fig.~\ref{fig:SF} we show the temperature dependence of $A^2$ closely resembles that of the superfluid density.

%
%



We note that one could also introduce $U(1)$  gauge fields into the BCS theory of dilute fermonic atom gases. In the fermion case, the gauge  fields couple to the diatom gap field, $\Delta$,  in much the same way as the  gauge field $W_\mu$ couples to the composite- (diatom-) field $A$ here.  Then one would have derived the fact that the superfluid density in BCS theory would be directly proportional to $\Delta^2$, as shown for Fermi systems in a more formal way from the Josephson relationship~\cite{r:josephson66} by Taylor~\cite{PhysRevB.77.144521}. 

Equation~\eqref{mass} is consistent with the Josephson relationship~\cite{r:josephson66}. This fact is apparent in region II, where the derivation of the Josephson relationship parallels the corresponding calculation performed by Taylor~\cite{PhysRevB.77.144521} for Fermi systems. 
In contrast with LOAF, where the field $A$ is composed of  bosons,  in BCS theory the gap, $\Delta$, is made up of fermions, causing a sign change in the loop contribution to the inverse  propagator. 
Following Taylor's methodology~\cite{PhysRevB.77.144521},  one obtains the superfluid density for a superfluid moving with velocity $\vec v$,  as
\begin{align}
\rho_s = \frac{1}{V} \left[ \frac{\partial^2 \Omega_v}{\partial v^2} \right]_{v=0}= - \frac { 8m^2 A^2 }{\hbar^2}  \lim_{q \rightarrow 0} \frac {1}{q^2 \, \tilde {\cal G}_{44}(q,0)} 
\>,
\label{josephA}
\end{align}
where   $\Omega_v = - T \ln Z_v$  is the grand canonical thermodynamic potential obtained by performing a twist operation~\cite{PhysRevA.8.1111} on the $A$ field, $A_v = e^{i2 m {\vec v} \cdot {\vec r}} A$, and $\tilde {\cal G}_{44}$ 
is the $A^\star A $ component of the composite field  propagator.
Just like in the BCS picture, in region~II,  $\rho_s \neq 0$ requires zero modes in the composite-field  propagators. 
Apart from a wave function renormalization, Eqs.~\eqref{mass} and~\eqref{josephA} are the same. 
We have shown recently~\cite{r:Dawson:2012} that the Josephson relation~\eqref{josephA} holds also in region~I, where $\phi \neq 0$ and $\chi^\prime = A$.


To summarize, in this paper we studied the physical properties of a new phase predicted by the LOAF approximation
to the dilute Bose gas theory. This phase resides in the phase diagram for temperatures between the critical temperature~$T_c$, where the BEC appears first, and the temperature~$T^{\star}$ corresponding to the onset of superfluidity.  We proved that this phase supports composite-field (diatom) Goldstone states accompanied by zero-energy and zero-momentum excitations in the system. 
The presence of superfluidity is tied to the existence of a U(1) charge-two diatom condensate in the system.
We also showed that the mass of the gauge particle generated through the Anderson-Higgs mechanism is related to the superfluid density via the Meissner effect. We find that 
the square of the anomalous density is a good approximation for the superfluid density. 

%
%
\begin{acknowledgments}
This work was performed in part under the auspices of the
U.S.~Department of Energy.   We would like to thank G.~Guralnik and A.~Saxena for useful discussions.
\end{acknowledgments}
%
%
\bibliography{johns}
%
%
\vfill
\end{document}